\documentclass{statsoc}
\usepackage[a4paper]{geometry}
\usepackage[utf8]{inputenc}
\usepackage[T1]{fontenc}

\usepackage{mathtools,amssymb,pifont}
\usepackage{graphicx,color,dsfont,multirow}
\usepackage{tabularx,pifont,booktabs,threeparttable}
\usepackage{bigdelim}   
\usepackage[most]{tcolorbox}
\usepackage{url}
\usepackage[caption=false]{subfig}

\usepackage{etoolbox}
\makeatletter
\patchcmd{\@makecaption}
  {\parbox}
  {\advance\@tempdima-\fontdimen2}
  {}{}
\makeatother

\usepackage{natbib}
\newcommand{\cmark}{\ding{51}}

\newcolumntype{L}[1]{>{\raggedright\arraybackslash}p{#1}}
\newcolumntype{C}[1]{>{\centering\arraybackslash}p{#1}}
\newcolumntype{R}[1]{>{\raggedleft\arraybackslash}p{#1}}  

\allowdisplaybreaks
    
\title[Leveraging Auxiliary Information in Nonignorable Models for Nonresponse]{Leveraging Auxiliary Information on Marginal Distributions in Nonignorable Models for Item and Unit Nonresponse}
\author[Akande {\it et al.}]{Olanrewaju Akande\footnote{\textit{Address for correspondence:} Social Science Research Institute, 140 Science Drive, Box 90989, Duke University, Durham, NC 27708 USA. \\ Email: olanrewaju.akande@duke.edu}, Gabriel Madson, D. Sunshine Hillygus and Jerome P. Reiter}
\address{Duke University, Durham NC, USA.}

\begin{document}
\begin{abstract}
Often, government agencies and survey organizations know the population counts or percentages for some of the variables in a survey.  These  may be available from auxiliary sources, for example, administrative databases or other high quality surveys. We present and illustrate a model-based framework for leveraging such auxiliary marginal information when handling unit and item nonresponse. We show how one can use the margins to specify different missingness mechanisms for each type of nonresponse. We use the framework to impute missing values in voter turnout in a subset of data from the U.S.\ Current Population Survey (CPS).  In doing so, we examine the sensitivity of results to different assumptions about the unit and item nonresponse.
\end{abstract}


\section{Introduction}
Many surveys have seen steep declines in response rates \citep{brick2013explaining, curtin2005changes}. Yet, government agencies and survey organizations---henceforth all called agencies---are under increasing budgetary pressures, making fewer resources available for extensive nonresponse follow-up activities. As a result, agencies are forced to account for missing values via statistical methods---for example, survey weight adjustments \citep{brickkalton} and variants of imputation \citep{andridgelittle,  kimfractional, rubin:1987}---that rely on strong assumptions about missing value mechanisms, e.g., all values are missing at random (MAR) \citep{rubin:1976}.  Such assumptions could be unrealistic, consequently threatening the validity and usefulness of inferences based on the survey data.

Agencies may be able to improve their procedures for dealing with missing values  by leveraging population-level information about the survey variables. For example, suppose a simple random sample has no unit nonresponse but has item nonresponse on the survey question asking the respondent's sex. If 70\% of participants report female, and the agency knows that the target population includes 50\% men and 50\% women, the agency likely should impute more men than women for the missing values in the survey question asking the respondent's sex.  In contrast, imputation routines that do not account for the known margin are likely to generate untrustworthy imputations. For example, a MAR model is likely to result in completed data with empirical percentages closer to 70\% female than 50\% female.   Of course, the agency should not use solely the population margin in imputations; it also should take advantage of observed information in other variables, so as to preserve multivariate relationships.

This example illustrates a broader context. An agency has access to accurate estimates of population percentages or counts for some variables in the survey.  These could be available from auxiliary data sources, such as censuses, administrative databases, high quality surveys, or  private sector data aggregators \citep{nrc08,nrc15}.  The agency seeks to take advantage of this auxiliary, population-level information in its methods for handling missing values, which could be due to both item and unit nonresponse.  In fact, agencies routinely find themselves in these scenarios; for example, many agencies use population counts as the basis for post-stratification adjustments for unit nonresponse. Usually, however, they do not use such margins in the imputation models for item nonresponse. 

In this article, we present and illustrate a model-based framework for leveraging auxiliary marginal information when handling both unit and item nonresponse. The margins allow agencies to weaken the assumptions about the reasons for missingness, while also offering 
flexibility in specification of missing data models.  In particular, we show how one can use the margins to specify different missingness mechanisms for unit and item nonresponse; for example, use a nonignorable model for unit nonresponse and an ignorable model for the variables with item nonresponse, or vice versa.
We apply the framework to handle missing values in a question on voter turnout in a subset of data from the U.S.\ Current Population Survey (CPS). Here, we know the actual number of voters in the election from published state-wide totals.  We use these auxiliary totals to generate model-based estimates of voter turnout, which we use for substantive empirical analyses examining voter turnout among various population subgroups (age, sex, and state).

Our work builds on the results of \citet{SadinleReiter2019}, who show how one can use auxiliary marginal information to specify identifiable, nonignorable models for data with item nonresponse. \citet{SadinleReiter2019} do not consider how to use auxiliary information to specify and estimate nonignorable models for data containing both item and unit nonresponse, which is a primary contribution of our work. 
We offer practical guidance on model specification, focusing on how analysts can use the auxiliary margins when modeling the unit and item nonresponse indicators. Via the CPS data analysis, we demonstrate processes for model evaluation and sensitivity analysis under different assumptions for unit and item nonresponse mechanisms.

The remainder of this article is organized as follows. In Section \ref{MD-AM}, we present the framework for specifying models for both unit and item nonresponse using information obtained from auxiliary sources, which we refer to as the MD-AM (missing data with auxiliary margins) framework \citep{Akande2019}. The MD-AM framework guides the specification of sequential additive nonignorable models \citep{HiranoEtAl1998, HiranoEtAl2001, SadinleReiter2019}, which we estimate by adapting the data augmentation strategy of \citet{schifeling:reiter}; we review these here. In Section \ref{illustration}, we present an application of the MD-AM framework using the CPS voter turnout data.  In Section \ref{discussion}, we conclude and discuss extensions of the MD-AM framework.

\section{The MD-AM Framework} \label{MD-AM}
The MD-AM framework is based on a two-step process for specifying a joint distribution for the survey variables and indicator variables for nonresponse. Specifically, we characterize the joint distribution using a sequential factorization of conditional models. We use the auxiliary information to guide the specification of the conditional distributions, using models that encode potentially nonignorable nonresponse mechanisms.  We require the models to be identifiable, which corresponds to the usual notion that any set of model parameter values maps to a unique value of the likelihood function (and vice versa).   Broadly, the two steps are as follows.

\textbf{{Step 1: Specify model for the observed data.}}
We begin with a model for the survey variables and nonresponse indicators that can be identifiable using the observed data alone, without any auxiliary information. Preferably, the model should allow for the maximum number of parameters identifiable from the observed data alone.  This can be done using either a selection model or a pattern mixture factorization \citep{GlynnEtAl1986,littlepatmix}, according to the analyst's preference. Generally, this step results in models, such as MAR models, that are often default choices for handling nonresponse in the missing data literature, absent auxiliary data. 

\textbf{{Step 2: Incorporate auxiliary margins.}}
We next find sets of parameters that can be added to the model in Step 1, so that the model still can be identified because of the auxiliary information.   Typically, there are multiple identifiable models---determined by the nature of the auxiliary information---each representing different assumptions about the missingness process. Agencies can choose from among these models according to interpretability and plausibility for the data at hand.  

In what follows, we demonstrate how to instantiate the MD-AM framework.  To facilitate understanding, we begin with a detailed look at a scenario involving two binary variables with both subject to item nonresponse, and where the data also suffer from unit nonresponse.  We use this scenario to show how the framework allows one to encode different missingness assumptions about unit and item nonresponse. In Section \ref{MD-AM:extension}, we discuss extensions to more variables and provide a general format for MD-AM models.

\subsection{Notation for the general MD-AM framework} \label{MD-AM:notation}
We first present notation for the MD-AM framework for general data scenarios; we adapt this notation for the illustrative scenario in Section \ref{scenariothree} and the general format for MD-AM models in Section \ref{MD-AM:extension}. Let $\mathcal{D}$ comprise data from the survey of $i = 1, \ldots, n$ individuals, and let $\mathcal{A}$ comprise data from the auxiliary database. Let $X = (X_1, \ldots, X_p)$ represent the $p$ variables in both $\mathcal{A}$ and $\mathcal{D}$, where each $X_k = (X_{1k},\ldots,X_{nk})^T$ for $k = 1, \ldots, p$.  Let $Y = (Y_{1}, \ldots, Y_q)$ represent the $q$ variables in $\mathcal{D}$ but not in $\mathcal{A}$, where each $Y_k = (Y_{1k},\ldots,Y_{nk})^T$ for $k = 1, \ldots, q$. We disregard variables in $\mathcal{A}$ but not $\mathcal{D}$ as they are not of primary interest.  We assume that $\mathcal{A}$ contains sets of marginal probabilities or counts for variables in $X$, summarized from some external database.

For each $k = 1, \ldots, p$, let $R^x_{ik} = 1$ if individual $i$ would not respond to the question on $X_k$ in the survey (i.e., $\mathcal{D}$), and $R^x_{ik} = 0$ otherwise. Similarly, for each $k = 1, \ldots, q$, let $R^y_{ik} = 1$ if individual $i$ would not respond to the question on $Y_k$ in the survey, and $R^y_{ik} = 0$ otherwise. Let $R^x = (R^x_1, \ldots, R^x_p)$ and $R^y = (R^y_1, \ldots, R^y_q)$, where each $R^x_k = (R^x_{1k}, \ldots, R^x_{nk})^T$ and $R^y_k = (R^y_{1k}, \ldots, R^y_{nk})^T$.  Let $R^x_i = (R^x_{i1}, \ldots, R^x_{ip})$ and $R^y_i = (R^y_{i1}, \ldots, R^y_{iq})$.
Let $U = (U_1, \ldots, U_n)$, where each $U_i = 1$ if individual $i$ would not respond to the survey at all (unit nonresponse), and $U_i = 0$ otherwise. We note that $(R^x_i, R^y_i)$ is observed for all cases $i$ with $U_i=0$, whereas $(R^x_i, R^y_i)$ is not observed for all cases $i$ with $U_i=1$.
Let $\mathcal{R} = (R^x, R^y)$. Finally, we define the observed data as $\mathcal{D}^{obs}=(\mathcal{D}, \mathcal{R}, U)$.

For simplicity, we use generic notations such as $f$ and $\eta$ for technically different functions and parameters respectively, but their actual meanings within each context should be clear within each context. 
For example, $f$, $\eta_{0}$, and $\eta_{1}$ need not be the same in the conditional probability mass functions $\Pr(X_1= 1|Y_1) = f(\eta_{0} + \eta_{1}Y_1)$ and $\Pr(Y_1 = 1 | X_1) = f(\eta_{0} + \eta_{1}X_1)$.

\subsection{MD-AM framework for illustrative scenario} \label{scenariothree}
Let $X=(X_1, X_2)$ comprise $p=2$ binary variables and $Y$ be empty, so that $\mathcal{D} = (X_1, X_2)$. Let $\mathcal{A}$ comprise the true marginal probabilities for $X_1$ and $X_2$ separately. We suppose that $X_1$ and $X_2$ suffer from item nonresponse, and some units do not respond at all to the survey.  Table \ref{scenariothree:marg} represents the relevant information in $\mathcal{D}^{obs}$ and $\mathcal{A}$ in a graphical format. 
\begin{table}
	\caption{\label{scenariothree:marg}Data structure for example with unit nonresponse and two variables subject to item nonresponse. Here, ``\cmark'' represents  observed components and ``\textbf{?}'' represents missing components.}
	\footnotesize
	\centering
	\begin{tabular}{r|c|c|c|c|c|} \cline{2-6}
			& $X_1$  & $X_2$ & $R^x_1$ & $R^x_2$ & $U$ \\ \cline{2-6}
			\multirow{5}{*}{$\mathcal{D}^{obs}$} \ldelim\{{5}{2.5mm} & \cmark & \cmark & 0 & \multirow{2}{*}{0} & \multirow{4}{*}{0} \\ \cline{2-4}
			& \textbf{?} & \cmark & 1 &  &  \\ \cline{2-5}
			& \cmark & \textbf{?} & 0 & \multirow{2}{*}{1} &  \\ \cline{2-4}
			& \textbf{?} & \textbf{?} & 1 &  & \\ \cline{2-6}
			& \textbf{?} & \textbf{?} & \textbf{?} & \textbf{?} & 1 \\ \cline{2-6}
			Auxiliary margin $\rightarrow$ & \cmark & \textbf{?} & \textbf{?} & \textbf{?} & \textbf{?} \\ \cline{2-6}
			Auxiliary margin $\rightarrow$ & \textbf{?} & \cmark & \textbf{?} & \textbf{?} & \textbf{?} \\ \cline{2-6}
	\end{tabular}
\end{table}

As a preliminary  step in the MD-AM model specification, we determine the number of identifiable parameters for the models in Step 1 and Step 2. To do so, it is useful to factor the joint distribution of $(X_2, X_1, R^x_2, R^x_1, U)$ into the product of $\theta_{xr_2r_1u} = \Pr(X_2=1|X_1=x,R^x_2=r_2,R^x_1=r_1,U=u)$, $\pi_{r_2r_1u} = \Pr(X_1=1|R^x_2=r_2,R^x_1=r_1,U=u)$,  $q_{r_1u} = \Pr(R^x_2=1|R^x_1=r_1,U=u)$, $s_{u} = \Pr(R^x_1=1|U=u)$ and $p = \Pr(U=1)$. We can estimate eight of these probabilities, namely ($p, s_0, q_{00}, q_{10}, \pi_{000}, \pi_{100}, \theta_{0000}, \theta_{1000}$), directly from $\mathcal{D}^{obs}$ alone.  Thus, we can identify eight parameters when specifying the models in Step 1 of the MD-AM framework.  The auxiliary margins $\Pr(X_1)$ and $\Pr(X_2)$ add two more pieces of information, which take the form of two constraints on the inestimable probabilities; see the online supplement for the constraints. Thus,  we can identify two additional parameters in Step 2 of the MD-AM framework.

We first follow Step 1 to specify a model for $\mathcal{D}^{obs}$ without using $\mathcal{A}$.  Here, and throughout this article, we use a selection model factorization, in which we first posit a model for $\mathcal{D}$, then models for $(U | \mathcal{D})$, and $(\mathcal{R} | \mathcal{D}, U)$. For this example in particular, we write the density $h(\mathcal{R} | \mathcal{D}, U) = h_2(R^x_2 | \mathcal{D}, U, R^x_1) h_1(R^x_1 | \mathcal{D}, U)$.  We present pattern mixture model factorizations in the online supplement.  A reasonable specification for Step 1 includes all parameters except those targeting the direct relationship between $R^x_1$ and $X_1$, the relationship between $R^x_2$ and $X_2$, and the relationships between $U$ and any of the other variables. As evident in the portion of Table \ref{scenariothree:marg} for $\mathcal{D}^{obs}$, none of these combinations of variables are fully observed together; hence, parameters that depend on simultaneous observations of them are not identifiable using $\mathcal{D}^{obs}$ alone.

Following this logic, we write the joint distribution of $(X_1,X_2)$ generically as 
\begin{align}
	\phantom{(X_1,X_2) \sim}
	&\begin{aligned} \label{scenariothree:selection1}
    \mathllap{(X_1,X_2) \sim} & \ f(X_1,X_2 | \Theta).
    \end{aligned}
\end{align}
In our example, we let $f$ be a fully saturated multinomial distribution.  In applications where the dimension of $X$ is larger, one can use other distributions, such as log-linear models or products of conditional regression models.
A default choice for the nonresponse indicator models is 
\begin{align}
	\phantom{\Pr(R^x_2 = 1 | X_1,X_2,U, R^x_1) =}
	&\begin{aligned} \label{scenariothree:selection2}
	\mathllap{\Pr(U = 1 | X_1, X_2) =} & \ g(\eta_{0})
	\end{aligned}\\
	&\begin{aligned} \label{scenariothree:selection3}
	\mathllap{\Pr(R^x_1 = 1 | X_1, X_2,U) =} & \ h_1(\zeta_{0} + \zeta_{1} X_2)
	\end{aligned}\\
	&\begin{aligned} \label{scenariothree:selection4}
	\mathllap{\Pr(R^x_2 = 1 | X_1, X_2,U, R^x_1) =} & \ h_2(\gamma_{0} + \gamma_{1} X_1).
	\end{aligned}
\end{align}
With $f$ as the fully saturated multinomial model, the combined model in \eqref{scenariothree:selection1}--\eqref{scenariothree:selection4} has eight free parameters, which is the maximum identifiable from $\mathcal{D}^{obs}$ alone.

We refer to this specification as MCAR+ICIN, as it assumes a missing completely at random (MCAR, \citet{littlerubin}) mechanism on the unit nonresponse indicator and versions of itemwise conditionally independent (ICIN) mechanisms \citep{sadinle:reiter:bmka} on both item nonresponse indicators. Generally, a missingness mechanism for any survey variable is said to follow the ICIN mechanism when the variable itself is conditionally independent of its missingness indicator, given the remaining survey variables and their missingness indicators. We note that ICIN encodes a nonignorable missingness mechanism since the remaining survey variables themselves may suffer from nonresponse, as is the case here.

In (\ref{scenariothree:selection2}) -- (\ref{scenariothree:selection4}), $\eta_{0}$ is identifiable since the marginal probability of $U$ is known from the observed data alone; $\zeta_{0}$ and $\zeta_{1}$ are identifiable since the joint relationship between $R^x_1$ and $X_2$ can be estimated from the observed data; and, $\gamma_{0}$ and $\gamma_{1}$ are identifiable since the joint relationship between $R^x_2$ and $X_1$ can be estimated from the observed data.

We next follow Step 2 of the MD-AM framework to incorporate the auxiliary information about $X_1$ and $X_2$. That is, we leverage the two additional constraints from $\mathcal{A}$ and relax some of the assumptions in the MCAR+ICIN model.  We can add two parameters to the models in multiple ways, reflecting different assumptions about the missingness mechanisms. We present three of them here. 

One option is to use the margins to enhance the models for item nonresponse, leaving the unit nonresponse as MCAR.  In particular, we couple (\ref{scenariothree:selection2}) with additive nonignorable (AN) models \citep{HiranoEtAl1998, HiranoEtAl2001} for $R_1^x$ and $R_2^x$, 
\begin{align}
	\phantom{\Pr(R^x_2 = 1 | X_1,X_2,U, R^x_1) =}
	&\begin{aligned} \label{scenariothree:selection6}
	\mathllap{\Pr(R^x_1 = 1 | X_1,X_2,U) =} & \ h_1(\zeta_{0} + \zeta_{1} X_2 + \zeta_{2} X_1)
	\end{aligned}\\
	&\begin{aligned} \label{scenariothree:selection7}
	\mathllap{\Pr(R^x_2 = 1 | X_1,X_2,U, R^x_1) =} & \ h_2(\gamma_{0} + \gamma_{1} X_1 + \gamma_{2} X_2).
	\end{aligned}
\end{align}
AN models have been used previously to handle attrition in longitudinal studies with refreshment samples \citep[e.g.,][]{nevo2003using,bhattacharya2008inference,DasEtAl2011,DengEtAl2013,schifelingcheng}, but not when handling unit and item nonresponse simultaneously as we do here.  AN models encode ignorable and nonignorable models as special cases. For example,  $(\zeta_{1} = 0, \zeta_{2} = 0)$ results in an MCAR mechanism for $R_1^x$; $(\zeta_{1} \neq 0, \zeta_{2} = 0)$ results in a conditionally MAR  mechanism for $R_1^x$; and, $\zeta_{2} \neq 0$ results in a missing not at random (MNAR, \citet{littlerubin}) mechanism for $R_1^x$.  Thus, the AN model uses a weaker assumption about the missingness than its special case models.  Importantly, while the AN model offers additional flexibility for modeling missingness, it is not assumption free---missing data always force one to make identifying assumptions.  For example, the AN model for $R_1^x$  posits that the reason for item nonresponse in $X_1$ depends on $X_1$ and $X_2$ through a function that is additive in $X_1$ and $X_2$.

As a general strategy, leveraging the marginal information for item nonresponse modeling is most appropriate when agencies consider item nonresponse to be potentially nonignorable and  unit nonresponse to be MCAR. It also can be preferred when agencies want to dedicate the auxiliary information to richer modeling of $R^x$ than $U$.  Agencies can do so when item nonresponse is a greater threat to the quality of inferences than unit nonresponse, for example, when the numbers of missing items for individual survey variables are larger than the number of unit nonrespondents.  For the illustrative scenario, we refer to this specification as AN-R.

Another option is to use the margins to enhance the model for unit nonresponse.  In particular, we couple (\ref{scenariothree:selection3}) and (\ref{scenariothree:selection4}) with an AN model for $U$, 
\begin{align}
	\phantom{\Pr(U = 1 | X_1,X_2) =}
	&\begin{aligned} \label{scenariothree:selection5}
	\mathllap{\Pr(U = 1 | X_1,X_2) =} & \ g(\eta_{0} + \eta_{1} X_1 + \eta_{2} X_2).
	\end{aligned}
\end{align}
The model in (\ref{scenariothree:selection5}) inherits the flexibility of AN models in the specification for the unit nonresponse model, in that it encompasses conditionally ignorable and nonignorable models as special cases. The model implies that the item nonresponse models are the same for unit respondents and nonrespondents.  We have to make this assumption, as we never observe $(R^x_{i1}, R^x_{i2})$ for cases with $U_i=1$.

Leveraging the marginal information for unit  nonresponse modeling is most appropriate when agencies consider unit nonresponse to be potentially nonignorable and the item nonresponse to be ICIN. It is appealing for situations where agencies feel it important to use richer models for $U$ than $R$, for example, when the amount of unit nonresponse is larger than the amount of item nonresponse.  For the illustrative scenario, we refer to this specification as AN-U.

As a final example, the MD-AM framework also encompasses a compromise between AN-R and AN-U.  We can use an ICIN model for unit nonresponse and one of the item nonresponse indicators, plus an AN model for the other item nonresponse indicator.  Specifically, we can use 
\begin{align}
	\phantom{\Pr(U = 1 | X_1,X_2) =}
	&\begin{aligned} \label{scenariothree:selection8}
	\mathllap{\Pr(U = 1 | X_1,X_2) =} & \ g(\eta_{0} + \eta_{2} X_2)
	\end{aligned}
\end{align}
coupled with (\ref{scenariothree:selection4}) and (\ref{scenariothree:selection6}), which we refer to as AN-R$^x_1$ in the simulations in the online supplement. Similarly, we can use 
\begin{align}
	\phantom{\Pr(U = 1 | X_1,X_2) =}
	&\begin{aligned} \label{scenariothree:selection9}
	\mathllap{\Pr(U = 1 | X_1,X_2) =} & \ g(\eta_{0} + \eta_{1} X_1)
	\end{aligned}
\end{align}
coupled with (\ref{scenariothree:selection3}) and (\ref{scenariothree:selection7}), which we refer to as AN-R$^x_2$ in the online supplement. Such models can be useful when $\mathcal{D}$ has a large amount of unit nonresponse, and only one of the variables has a large amount of item nonresponse (and the other does not).  In this way, we utilize the information from $\mathcal{A}$ to enrich the models for both unit and item nonresponse.

In the online supplement, we illustrate the different specifications empirically using simulation studies. An important takeaway from the studies, as well as the three specifications presented here, is that we generally cannot specify AN models for the unit nonresponse indicator and all item nonresponse indicators simultaneously since such models cannot be identified from the information available in the data. It is thus important that agencies use substantive knowledge to inform model specifications and analyze the sensitivity of results to multiple plausible model specifications, as explained below.

\subsection{Implementation considerations}\label{MD-AM:extension}
The example in Section \ref{scenariothree} illustrates how the MD-AM framework enables agencies to tailor their use of information in the auxiliary marginal distributions. For example, agencies can use $\mathcal{A}$ to specify AN models for unit (item) nonresponse indicator when they want to dedicate model flexibility for unit (item) nonresponse models.  However, agencies need not select only one model in the MD-AM framework.  They can examine sensitivity of inferences to different specifications, as we do in our CPS application and also implicitly in the simulations in the online supplement. When using the MD-AM framework to release multiple imputations under different assumptions about the missingness, agencies can use the approach of \citet{siddique:hrel:crespi}, to incorporate uncertainty regarding the missing data mechanism.

Extending the MD-AM approach to more variables, as well as categorical variables with more than two levels, is conceptually straightforward. When the data include $Y$ variables with missing values, as is usually the case, we simply add $Y$ in the models for $f(\mathcal{D})$ and add conditional models for each $R^y_k$. We recommend putting the conditional models for the $R^y_k$'s at the end of the sequence.  Because we do not have marginal distributions for these $Y$ variables, we are forced to make stronger assumptions about them, such as ICIN or MAR.  When the data include multiple variables with margins in $\mathcal{A}$, we recommend treating them as we do $X_1$ and $X_2$ in Section \ref{scenariothree}.  For example, if we have the marginal distribution for some variable $X_k$, we can add terms involving $X_k$ to the item nonresponse model for $R^x_k$ or to the unit nonresponse model for $U$.  As before, with the MD-AM framework, the agency can choose where to dedicate the extra modeling flexibility.

Mathematically, we define a general version of the MD-AM framework using a sequence of conditional models.  First, we specify some model for the joint distribution of $\mathcal{D}=(X, Y)$ given parameters $\Theta$ generically as 
\begin{align}
	\phantom{\mathcal{D} \sim}
	&\begin{aligned} \label{general:XandY}
    \mathllap{\mathcal{D} \sim} & \ f(\mathcal{D} | \Theta).
    \end{aligned}
\end{align}
Second, we specify models for the unit nonresponse indicator and item nonresponse indicators.
Let $M_U(\mathcal{D}, \Omega_0)$ represent a linear predictor for regression modeling, that is, a function of $\mathcal{D}$ that is linear in parameters $\Omega_0$.  We require$M_U(\mathcal{D}, \Omega_0)$ to result in an identifiable model for $U$ in Step 1, without using $\mathcal{A}$, such as a MCAR or an ICIN model. Similarly, let $M_{R^x_k}(\mathcal{D},  \Phi_{0k})$ and $M_{R^y_k}(\mathcal{D}, \Psi_{0k})$ represent linear predictors with parameters $\Phi_{0k}$ and $\Psi_{0k}$, respectively, that  result in identifiable models for each ${R^x_k}$ and ${R^y_k}$, respectively, in Step 1 without using $\mathcal{A}$.
Let $\mathcal{U}$ be the set of all $X_k$'s chosen to be in the model for $U$.  We write the nonresponse indicator models in MD-AM for $U$, any arbitrary $R_k^x$, and any arbitrary $R_k^y$ as 
\begin{align}
	\phantom{\Pr(R^x_2 = 1 | X_1,X_2,U, R^x_1) =}
	&\begin{aligned} \label{general:U}
	\mathllap{\Pr(U = 1 | X, Y) =} & \ g ( M_U(\mathcal{D}, \Omega_0) + \sum_{k=1}^p f_k(X_k, \Omega_{1k}) I(X_k \in \mathcal{U}))
	\end{aligned}\\
	&\begin{aligned} \label{general:R1}
	\mathllap{\Pr(R^x_k = 1 | X, Y, U)  =} & \ h_k ( M_{R^x_k}(\mathcal{D}, \Phi_{0k}) + f_k(X_k, \Phi_{1k}) I(X_k \notin \mathcal{U}) )
	\end{aligned}\\
	&\begin{aligned} \label{general:Rp}
	\mathllap{\Pr(R^y_k = 1 | X, Y, U, R^x)  =} & \ s_k ( M_{R^y_k}(\mathcal{D}, \Psi_{0k}) ).
	\end{aligned}
\end{align}
Here, $g$, $h_k$ and $s_k$ are mean functions or inverse link functions dependent on the model specification; $f_k(X_k, \Gamma_k )$ is a function that maps the categorical variable $X_k$ onto a combination of dummy variables for its levels, which are linear in the set of parameters $\Gamma_k$; and, $I(\cdot)=1$  when its argument is true and $I(\cdot)=0$ otherwise.

When specifying the sequence of conditional models, one needs to decide the ordering of the variables. For the survey variables, the order is somewhat arbitrary, in that we seek to characterize their joint distribution.  For practicality, we recommend following the advice in typical missing data imputation routines \citep{burgreit10,vanbuuren:06,vanbuuren,Buuren2012} and ordering from least to most missing values.  For the nonresponse indicators, we find it convenient to put variables with auxiliary margins early in the sequence and variables without auxiliary margins later in the sequence.  It can be easier to interpret the nonresponse mechanisms, and thus decide how to use the information in $\mathcal{A}$, in models with fewer terms, which is the case for the models early in the sequence.  In our simulations, the ordering of the nonresponse indicators does not seem to affect the results noticeably, especially when we do not use the nonresponse indicators as predictors. Nonetheless, agencies can assess sensitivity of results to orderings of the variables.

Theoretically, it might be impossible to distinguish between unit nonresponse and item nonresponse for some cases, e.g., when an individual does not provide information on any of the questions used in a particular analysis.  This is not a problem in all our simulation scenarios by design. However, when agencies cannot distinguish the two forms of nonresponse, they may need to incorporate assumptions about the nonresponse indicators into the modeling in order to have identifiable models. For example, agencies can treat individuals who do not respond to any of the questions being analyzed as unit nonrespondents. In this case, we add the constraint of zero probability to the chance that all item nonresponse indicators equal one; for example, set $\Pr(R^x_2 = 1)$ to zero whenever $R^x_1 = 1$ and vice versa.  In this way, the model for $U$ completely captures all unit nonrespondents plus item nonrespondents who do not respond to any questions.

\section{Application to CPS Voter Turnout Data} \label{illustration}
Voter turnout is the cornerstone of electoral democracy. Yet, voter turnout in the United States is considered among the worst in advanced democracies, with less than half the population casting a ballot in recent presidential elections \citep{mcdonald2001myth}. Turnout in the United States also varies dramatically across different demographic and geographic subgroups in the population, shaping not only who gets elected but also what policies get implemented \citep{LeighleyNagler2013}. Although there is widespread recognition of low and unequal electoral participation, data limitations have impeded a better understanding of civic participation in American democracy.

Given that we have official government counts of ballots cast in an election, it might seem puzzling that calculating turnout rates is at all complicated.  There are two problems. First, we often lack demographic information in administrative election records. Although states maintain voter registration records that indicate if a resident voted in an election, there is considerable variation in the demographic data collected in those files.  For example, not all states collect information about race and ethnicity.  Second, to calculate a turnout rate, we need an estimate of the denominator---the voting eligible population (VEP), rather than the total population \citep{mcdonald2001myth}. Given these issues, researchers typically must rely on survey-based estimates of turnout rather than administrative data.

Among surveys, the CPS is considered the gold standard for estimating voter turnout. Every Congressional election year the CPS November Supplement asks a variety of questions about voter registration and turnout to U.S. citizen adults in sampled households. As a result, the CPS is one of the few surveys with sufficient sample size to make turnout estimates by state, as the sample size exceeds 75,000 voting-age citizens, stratified by state. Nonetheless, the CPS voter turnout measure is plagued by high levels of missing data, as we document in Section \ref{data}.   As we show in Section \ref{results}, this nonresponse appears to be nonignorable.  Thus, we apply the MD-AM framework to data from the CPS, leveraging auxiliary information to estimate voter turnout for demographic and geographic subgroups of the population.

\subsection{Data}\label{data}
We analyze data from the 2012 CPS November supplement. We obtain person-level data from the Integrated Public Use Microdata Series (IPUMS). We focus our analysis on four states: Florida (FL), Georgia (GA), North Carolina (NC) and South Carolina (SC). All are southern states that vary somewhat in demographic composition, as well as their battleground election status in 2012. We use the four variables described in Table \ref{variable:description}.
\begin{table}
	\caption{\label{variable:description}Description of variables used in illustration.}
	\footnotesize
	\centering
	\begin{tabular}[c]{ll}
		\toprule
		Variable    & Categories \\
		\midrule
		State & 1 = Florida, 2 = Georgia, 3 = North Carolina, 4 = South Carolina \\
		Sex & 0 = Male, 1 = Female \\
		Age & 1 = 18 - 29, 2 = 30 - 49, 3 = 50 - 69, 4 = 70+ \\
		Vote & 0 = Did not vote; 1 = Voted \\
		\bottomrule
	\end{tabular}
\end{table}
The resulting dataset comprises $n = 11,846$ individuals (5,086 in FL; 2,475 in GA; 2,519 in NC; and 1,766 in SC). Missing data rates are reported in Table \ref{nonresponserates}.
\begin{table}
	\caption{\label{nonresponserates}Person-level unit and item nonresponse rates by state in the CPS data.  Only 7 total cases (six in FL and 1 in SC) are missing sex.}
	\footnotesize
	\centering
	\begin{tabular}{lcccc}
		\toprule
		& \multirow{2}{*}{Unit} & \multicolumn{3}{c}{Item} \\ \cmidrule(lr){3-5}
		& & Vote & Sex & Age \\
		\midrule
		FL & .28 & .18 & .00 & .07 \\
		GA & .21 & .16 & .00 & .05 \\
		NC & .24 & .11 & .00 & .03 \\
		SC & .25 & .10 & .00 & .03 \\
		\bottomrule
	\end{tabular}
\end{table}

Across states, item nonresponse is substantial for the vote question, low for age, and trivial on the sex variable.  A CPS respondent is flagged for item nonresponse on the vote question if they are adult citizens in a responding CPS household who refused to answer the voting question (``Refused'': 1.8\%), did know know the answer to the voting question (``Don't Know'': 1.4\%), or were never asked the voting question (``No Response'': 11.5\%). Some of this missing data on vote is attributable to proxy responding, where one individual in a household answers on behalf of all members of the household but does not know if other household members voted.  For example, 87.5\% of those answering ``Don't Know'' to the voting question were proxy respondents. 
The sizeable fraction of ``No Response'' is likely due to respondents not being asked the vote question if their age or citizenship is missing at the time the voter supplement was in the field.  For some participants, the Census Bureau determines values for the missing age and citizenship subsequent to data collection.
The Census Bureau uses hot deck imputation for age and sex (but not voting) in the released data file, so we code age and sex as missing if the variable was flagged as being edited in the data file.

The CPS does not report unit nonresponse rates for persons, as the CPS is sampled at the household-level. We used the following steps to estimate the number of individuals eligible for the November supplement (U.S. citizens of voting age) among nonresponding households. In the household sampling data file, the Census Bureau provides disposition information on all sampled households that failed to respond.  After excluding households deemed ineligible for the CPS survey (labelled ``Type C'' in the data file), we estimate the average number of adult citizens per household in each state---estimated from the Census Bureau's special tabulation of the Citizen Voting Age Population (CVAP) using 5-year American Community Survey (ACS) data---divided by the total number of housing units in those states.  We multiply this average by the number of eligible nonresponding households in the CPS, resulting in individual-level estimates of unit nonresponse, rounded to the nearest person.
These numbers are used to derive the unit nonresponse rates in Table \ref{nonresponserates}.  In the resulting data file, these cases have no information beyond state of residence. 

As auxiliary marginal information for the MD-AM models, we use the state's voter turnout rate calculated from the official number of ballots cast for the highest office on the ballot divided by estimates of VEP, provided by The United States Elections Project (USEP)(\url{http://www.electproject.org/2012g}), which compiles government data to create election year estimates of the VEP from the American Community Survey and Department of Justice felon estimates \citep{mcdonald2008united}.  These percentages are as follows:  FL = 62.8\%, GA = 59.0\%, NC = 64.8 \% and SC = 56.3\%. We also use marginal information for the age distribution in each state from the 2010 census, displayed in Table \ref{demographics}.  Although slightly older than the 2012 CPS, the decennial census offers the advantage of providing the most accurate data on demographic characteristics of the population of each state.
\begin{table}
\caption{\label{demographics}Population-level margins for age groups by state. For comparison, entries in ``Available Cases'' are based on all available cases in the CPS data for the age groups.}
\centering
\begin{tabular}{rccccccccc}
  \toprule
  & \multicolumn{4}{c}{\textbf{Margin}} & & \multicolumn{4}{c}{\textbf{Available Cases}} \\
 \cmidrule(lr){2-5} \cmidrule(lr){7-10}
 & FL & GA & NC & SC & & FL & GA & NC & SC \\ 
  \midrule
  $<$30 & .20 & .23 & .22 & .22 & & .16 & .20 & .17 & .17  \\ 
  30-49 & .33 & .39 & .37 & .34 & & .30 & .39 & .34 & .32  \\ 
  50-69 & .31 & .29 & .30 & .32 & & .36 & .32 & .33 & .39  \\ 
  70+ & .16 & .09 & .11 & .12 & & .18 & .09 & .16 & .13  \\
   \bottomrule
\end{tabular}
\end{table}

\subsection{Modeling}
Let $S_i$, $G_i$, $A_i$ and $V_i$  represent the state, sex, age and vote of the $i=1, \dots, n$ individuals in the data. All four variables are in $\mathcal{D}$.  Population margins for all variables but $G_i$ are in $\mathcal{A}$. We do not rely on auxiliary margins for sex from the 2010 census, primarily to illustrate how the absence of auxiliary information affects model specification. We note that only seven survey participants are missing a value for sex.  As before, let $U_i$ represent the unit nonresponse indicator for individual $i$. Also, let $R^{y}_{iG}$, $R^{x}_{iA}$ and $R^{x}_{iV}$ be item nonresponse indicators for individual $i$, for sex, age and vote respectively, where each equals one if the corresponding variable is missing and equals zero otherwise.

The item nonresponse in all states has a monotone pattern:  age and vote are always missing when sex is missing, and vote is always missing when age is missing. We therefore include constraints in the item nonresponse models to respect the monotone patterns.  Without these constraints, the models would fail to capture the nonresponse process adequately.

The relationships between $S_i$ and the other variables, as well as the nonresponse indicators, are always observed. Therefore, we use $S_i$ as a covariate in all the models. Following Step 1 of the MD-AM framework, we specify the following models for the  distribution of $(G_i, A_i, V_i)$.
\begin{align}
\phantom{V_i | A_i, G_i, S_i\sim}
&\begin{aligned} \label{illustration:selection1}
\mathllap{G_i | S_i \sim} & \ Bern(\pi^{G}_i); \\
& \ logit(\pi^{G}_i) = \beta_{1} + \beta_{2,s} \mathds{1}[S_i=s] \\
\end{aligned}\\
&\begin{aligned} \label{illustration:selection2}
\mathllap{A_i | G_i, S_i \sim} & \ Cat(Pr[A_i \leq a] - Pr[A_i \leq a-1]); \\
& \ logit(Pr[A_i \leq a]) = \phi_{1,a} + \phi_{2,s} \mathds{1}[S_i=s] + \phi_{3} G_i
\end{aligned}\\
&\begin{aligned} \label{illustration:selection3}
\mathllap{V_i | A_i, G_i, S_i \sim} & \ Bern(\pi^{V}_i); \\
& \ logit(\pi^{V}_i) = \ \nu_{1} + \nu_{2,s} \mathds{1}[S_i=s] + \nu_{3,a} \mathds{1}[A_i=a] + \nu_{4} G_i \\
& \ \ \ \ \ \ \ \ \ \ \ \ \ \ \ \ \ + \nu_{5,sa} \mathds{1}[S_i=s, A_i=a].
\end{aligned}
\end{align}
Here, the model for $A_i$ is a proportional odds regression, and the models for $G_i$ and $V_i$ are logistic regressions. For parsimony, we exclude all interaction terms except $\mathds{1}[S_i=s, A_i=a]$. We do not see evidence that the excluded interaction terms have important predictive power based on exploratory data analysis. In scenarios where the interaction terms are potentially important or of inferential interest, they can be included since they can be estimated based on our discussions in Section \ref{MD-AM}. 

Following Step 2 of the MD-AM framework, we next specify models for the nonresponse indicators.  Since we have a high unit nonresponse rate and a low item nonresponse rate for age, and since Table \ref{demographics} indicates some differences between the age distributions for the CPS respondents and the population, we use the margin for $A$ to enrich the model for $U$ and sacrifice the use of an AN model for $R^{x}_A$. We therefore use an ICIN model for $R^{x}_A$, as a default option for estimating the maximum number of parameters possible based on the remaining information. Since we do not have a margin for sex, we cannot include $G$ in the models for $R^{y}_G$ or $U$. Since the nonresponse rate for sex is so low, we adopt a MAR model for $R^{y}_G$ for computational convenience, even though we can identify parameters for $A$ and $V$ in the model for $R^{y}_G$.

Since we are most interested in estimating turnout, and $V$ has a high item nonresponse rate in addition to being missing for unit nonrespondents, we consider two ways to leverage the population-level turnout rates. First, we use the margin for $V$ to estimate a nonignorable model for $R^{x}_V$. Second, we use the margin for $V$ to estimate a nonignorable model for $U$, dependent on $V$. We present the findings from both models in Section \ref{results}.

\subsubsection{Using the turnout margins in nonresponse models}  \label{sec:turnoutitem}
One might be concerned that item nonresponse to the vote question depends on whether or not the participant actually voted.  To write a model that incorporates this possibility, we write our specification of the selection models as
\begin{align}
\phantom{R^{x}_{iV}  | R^{A}_i, R^{G}_i, U_i,\ldots \sim}
&\begin{aligned} \label{illustration:selection4}
\mathllap{U_i  | \ldots \sim} & \ Bern(\phi^{U}_i); \\
& \ logit(\phi^{U}_i) = \gamma_{1} + \gamma_{2,s} \mathds{1}[S_i=s] + \gamma_{3,a} \mathds{1}[A_i=a]
\end{aligned}\\
&\begin{aligned} \label{illustration:selection5}
\mathllap{R^{y}_{iG}  | U_i,\ldots \sim} & \ Bern(\phi^{G}_i); \\
& \ logit(\phi^{G}_{i}) = \eta_{1} + \eta_{2,s} \mathds{1}[S_i=s]
\end{aligned}\\
&\begin{aligned} \label{illustration:selection6}
\mathllap{R^{x}_{iA} | R^{y}_{iG}, U_i,\ldots \sim} & \ Bern([\phi^{A}_{i}]^{(1-R^{y}_{iG})}); \\
& \ logit(\phi^{A}_{i}) = \alpha_{1} + \alpha_{2,s} \mathds{1}[S_i=s] + \alpha_{3} G_i   + \alpha_{4} V_i
\end{aligned}\\
&\begin{aligned} \label{illustration:selection7}
\mathllap{R^{x}_{iV}  | R^{x}_{iA}, R^{y}_{iG}, U_i,\ldots \sim} & \ Bern([\phi^{V}_{i}]^{(1-R^{x}_{iA})}); \\
& \ logit(\phi^{V}_{i}) = \psi_{1} + \psi_{2,s} \mathds{1}[S_i=s] + \psi_{3,a} \mathds{1}[A_i=a] \\
& \ \ \ \ \ \ \ \ \ \ \ \ \ \ \ \ \ \ + \psi_{4} G_i + \psi_{5} V_i.
\end{aligned}
\end{align}
In each model, ``$\ldots$'' represents conditioning on $(V_i, A_i, G_i, S_i)$. 
We cannot add $\gamma_{4} V_i$ to (\ref{illustration:selection4}), because this information can only come from the margin for vote ($V_i$ and $U_i$ are never observed together) and we have chosen to use the margin to estimate $\psi_{5}$ in (\ref{illustration:selection7}). We refer to the  model that uses $V$ in the model for $R^x_V$ as MD-R.

Given the substantial unit nonresponse, it is also reasonable to enrich the unit nonresponse model instead of the model for $R^x_V$. To do so, we modify (\ref{illustration:selection4}) and (\ref{illustration:selection7}) as follows.
\begin{align}
\phantom{R^{V}_i  | R^{A}_i, R^{G}_i, U_i,\ldots \sim}
&\begin{aligned} \label{illustration:selection8}
\mathllap{U_i  | \ldots \sim} & \ Bern(\phi^{U}_i); \\ 
& \ logit(\phi^U_{i}) = \gamma_{1} + \gamma_{2,s} \mathds{1}[S_i=s] + \gamma_{3,a} \mathds{1}[A_i=a] + \gamma_4 V_i
\end{aligned}\\
&\begin{aligned} \label{illustration:selection9}
\mathllap{R^{x}_{iV}  | R^{x}_{iA}, R^{y}_{iG}, U_i,\ldots \sim} & \ Bern([\phi^{V}_i]^{(1-R^{x}_{iA})}); \\ 
& \ logit(\phi^{V}_i) = \psi_{1} + \psi_{2,s} \mathds{1}[S_i=j] + \psi_{3,a} \mathds{1}[A_i=j] + \psi_{4} G_i.
\end{aligned}
\end{align}
We continue to use the item nonresponse models in \eqref{illustration:selection5} and \eqref{illustration:selection6}. We refer to the  model that uses $V$ in the model for $U$ as MD-U.

It is not possible to know for certain which one of these MD-AM models best describes the full data distribution. However, there are theoretical reasons to expect participation in an election and participation in a survey are strongly related (e.g., \citeauthor{brehm2009phantom} \citeyear{brehm2009phantom})---suggesting the MD-U model is more appropriate.  Previous research has similarly found that unit nonresponse among nonvoters can bias upward estimates of voter turnout \citep{burden2000voter}. Additionally, the unit nonresponse rates are higher than the item nonresponse rates for vote, suggesting that  we can benefit more from allocating the additional modeling flexibility to the model for $U$.

However, we need not choose only one of the models.  Rather, can use the two sets of results to portray the sensitivity of results to potentially nonignorable missing data, as we do in Section \ref{results}.

\subsubsection{Fitting the models}
For both MD-U and MD-R, to incorporate the auxiliary margins, we follow the approach of \citet{schifeling:reiter} by augmenting the observed data with a large number of synthetic observations with empirical distributions that match the marginal probabilities in $\mathcal{A}$. Specifically, we generate $n^\star = 3n$ synthetic observations for each of the variables with available auxiliary margins, resulting in a total of $71,076$ synthetic observations added to the observed data.  We leave values of $(U, R^{y}_G, R^{x}_A, R^{x}_V)$ completely missing for the synthetic observations. For each margin, we augment with three times the size of the observed data so that the empirical margins match the auxiliary information with negligible standard error.  Following \citet{schifeling:reiter}, we impute these missing values as part of a Bayesian MCMC sampler, using predictive distributions derived from the full model specification.

By using a large $n^\star$, we treat the auxiliary margins as having negligible standard errors. When $\mathcal{A}$ has non-negligible uncertainty, one can make $n^\star$ smaller to correspond to the desired standard error, following the approach in \citet{schifeling:reiter} and as explained in the online supplement.

We fit all models using Bayesian MCMC, with non-informative priors for all parameters. Details of the MCMC sampler are provided in the supplement.  We run the MCMC sampler for 10,000 iterations, discarding the first 5,000 as burn-in, resulting in 5,000 posterior samples. We base inferences on all 5,000 post burn-in posterior samples. Although we use posterior inference throughout this article, one also could use approximately independent draws from the posterior samples to perform multiple imputation (MI) \citep{CarpenterKenward2013, reiter:raghu:07, rubin:1987, schafer}, which can be appealing for data dissemination \citep{alanya:wolf:sotto, little:vart, peytchev12}.

\subsection{Results} \label{results}

We start by presenting the full set of results for MD-U and MD-R. Table \ref{results:cc} and Table \ref{results:cc:Carolinas} summarize the inferences for voter turnout rates for various demographic subgroups for both MD-AM models, along with point estimates from the complete cases.
\begin{table}
\caption{\label{results:cc}Turnout estimates, with corresponding 95\% credible intervals in parentheses, of sub-populations by state for MD-AM models in Florida and Georgia. MD-R uses  $V$ in the model for $R^x_V$, MD-U uses $V$ in the model for $U$, and CC is the results from the complete cases.  ``M'' stands for male and ``F'' stands for female. The population-level margins are 62.8\% in FL and 59.0\% in GA.}
\footnotesize
\centering
\begin{tabular}{rcccccc}
  \toprule
 & \multicolumn{3}{c}{\textbf{FL}} & \multicolumn{3}{c}{\textbf{GA}} \\
 \cmidrule(lr){2-4} \cmidrule(lr){5-7}
 & MD-R & MD-U & CC  & MD-R & MD-U & CC \\ 
  \midrule
  Full & .62 (.62, .63) & .62 (.61, .63) & .75 &  .59 (.58, .60) & .60 (.59, .61) & .73    \\ 
  M & .60 (.58, .61) & .59 (.57, .60) & .73  & .57 (.55, .58) & .56 (.54, .58) & .71  \\ 
  F & .64 (.63, .66) & .65 (.64, .67) & .77  & .62 (.60, .63) & .63 (.61, .64) & .75  \\ 
  $<$30 & .47 (.44, .50) & .38 (.35, .42) & .55 &  .44 (.40, .48) & .38 (.34, .42) & .56   \\ 
  30-49 & .60 (.57, .63) & .57 (.54, .60) & .73 &  .61 (.58, .64) &  .60 (.57, .64) & .74   \\ 
  50-69 & .69 (.66, .71) & .75 (.72, .78) & .82 &  .70 (.67, .73) & .76 (.72, .80) & .83   \\ 
  70+ & .72 (.69, .75) & .77 (.73, .82) & .84 &  .62 (.55, .69) & .65 (.56, .73) & .76   \\ 
  $<$30(M) & .45 (.41, .48) & .35 (.31, .39) & .52  & .41 (.37, .46) & .35 (.30, .39) & .49  \\ 
  30-49(M) & .58 (.55, .61) & .54 (.51, .57) & .69  & .58 (.55, .62) & .57 (.54, .61) & .70  \\ 
  50-69(M) & .67 (.64, .69) & .73 (.69, .75) & .80  & .68 (.64, .71) & .74 (.69, .78) & .83   \\ 
  70+(M) & .70 (.67, .74) & .75 (.70, .80) & .86 &  .59 (.52, .67) & .61 (.52, .70) & .81   \\ 
  $<$30(F) & .50 (.46, .53) & .41 (.37, .45) & .58  & .46 (.42, .51) & .41 (.37, .45) & .61   \\ 
  30-49(F) & .62 (.60, .65) & .60 (.57, .63) & .77 & .63 (.60, .66) & .63 (.60, .67) & .77   \\ 
  50-69(F) & .71 (.68, .73) & .77 (.74, .80) & .84  & .72 (.68, .75) & .78 (.75, .82) & .83   \\ 
  70+(F) & .74 (.70, .77) & .80 (.75, .84) & .81 & .64 (.57, .71) & .67 (.59, .75) & .73   \\ 
   \bottomrule
\end{tabular}
\end{table}
Across all states, estimates of the full state-wide turnout rates from both MD-AM models are close to the actual population-level turnout rates.  While there are some differences in the subgroup results of the two MD-AM models, which we return to shortly, both models reproduce well-documented patterns in voter turnout. For example, consistent with other surveys and official voting records, women vote with higher frequency than men \citep{LeighleyNagler2013}. Age is also strongly related to turnout, with young adults in the United States turning out at much lower rates than older citizens \citep{holbein2020making}. Likewise, young men have the lowest predicted turnout rates across all states in both models.
\begin{table}
\caption{\label{results:cc:Carolinas}Turnout estimates, with corresponding 95\% credible intervals in parentheses, of sub-populations by state for MD-AM models in North Carolina and South Carolina. MD-R uses  $V$ in the model for $R^x_V$, MD-U uses $V$ in the model for $U$, and CC is the results from the complete cases.  ``M'' stands for male and ``F'' stands for female. The population-level margins are 64.8\% in NC and 56.3\% in SC.}
\footnotesize
\centering
\begin{tabular}{rcccccc}
  \toprule
 & \multicolumn{3}{c}{\textbf{NC}} & \multicolumn{3}{c}{\textbf{SC}} \\
 \cmidrule(lr){2-4} \cmidrule(lr){5-7} 
 & MD-R & MD-U & CC & MD-R & MD-U & CC \\ 
  \midrule
  Full  & .65 (.64, .66) & .64 (.63, .65) & .77 & .57 (.56, .58) & .56 (.55, .57) & .73 \\ 
  M  & .63 (.61, .64) & .61 (.59, .62) & .76 & .54 (.53, .56) & .52 (.51, .54) & .68 \\ 
  F &  .67 (.66, .68) & .67 (.66, .69) & .79 & .59 (.58, .61) & .59 (.57, .61) & .77 \\ 
  $<$30  & .50 (.45, .56) & .45 (.40, .51) & .64 & .46 (.40, .52) & .42 (.36, .48) & .64 \\ 
  30-49 & .65 (.61, .68) & .62 (.58, .65) & .76 & .54 (.49, .58) & .51 (.47, .55) & .70 \\ 
  50-69  & .72 (.68, .75) & .75 (.71, .79) & .82 & .67 (.63, .71) & .69 (.65, .73) & .78 \\ 
  70+ & .76 (.71, .81) & .79 (.74, .85) & .84 & .61 (.53, .68) & .62 (.54, .72) & .75 \\ 
  $<$30(M) & .48 (.43, .53) & .41 (.37, .47) & .58 & .44 (.38, .49) & .39 (.33, .45) & .57 \\ 
  30-49(M) &  .63 (.59, .66) & .58 (.54, .62) & .72 & .51 (.47, .56) & .48 (.43, .52) & .67 \\ 
  50-69(M)  & .69 (.66, .73) & .72 (.68, .76) & .83 & .65 (.61, .69) & .66 (.61, .70) & .71 \\ 
  70+(M)  & .74 (.69, .79) & .77 (.71, .83) & .88 & .58 (.51, .66) & .59 (.50, .69) & .76 \\ 
  $<$30(F)  & .53 (.47, .58) & .48 (.43, .54) & .68 & .48 (.42, .54) & .45 (.39, .51) & .72 \\ 
  30-49(F)  & .67 (.63, .70) & .64 (.61, .68) & .81 & .56 (.52, .60) & .54 (.50, .59) & .73 \\ 
  50-69(F)  & .73 (.70, .77) & .77 (.73, .81) & .82 & .69 (.65, .73) & .71 (.67, .76) & .83 \\ 
  70+(F)  & .78 (.73, .82) & .81 (.76, .86) & .82 & .63 (.55, .70) & .65 (.57, .74) & .74 \\ 
   \bottomrule
\end{tabular} 
\end{table}

Notably, the estimates from the MD-AM models appear more accurate than the corresponding complete-case estimates. For example, the MD-AM models shrink the statewide turnout rate in Florida by around 13 percentage points, from an implausible 75\% to 62\%---close to official election estimates. More importantly, the MD-AM results for subgroups seem more plausible as well. For example, in Florida the complete case point estimates for all the age subgroups except those $<30$ exceed the population turnout rate of 62\%.  In contrast, 
by adjusting for nonignorable nonresponse, the MD-AM models shrink the complete case estimates for subgroups toward values more in line with administrative records.  The amount of the adjustment differs by subgroup; for example, in Georgia the MD-U models shrinks the point estimate of the turnout rate of young females (<30 years of age) by 20 percentage points---from an unrealistic complete case estimate of 61\% to a  believable point estimate of 41\%---and  shrinks the point estimate of turnout rate of young males by 14 percentage points. A similar pattern is seen with the MD-R model, although with different numbers.

The two MD-AM models produce similar inferences at the state level and for many of the subgroups; however, there are some differences. The subgroup point estimates from MD-U generally are lower for younger voters and higher for older voters than those from MD-R. Put another way, the MD-U model suggests a larger turnout gap between younger and older citizens than the MD-R model. Many of the corresponding 95\% credible intervals overlap, however.

\begin{figure}
\centering
\subfloat[Predicted turnout among item nonrespondents for MD-R model.]{
\makebox{\includegraphics[width=0.45\linewidth,height=3.7cm]{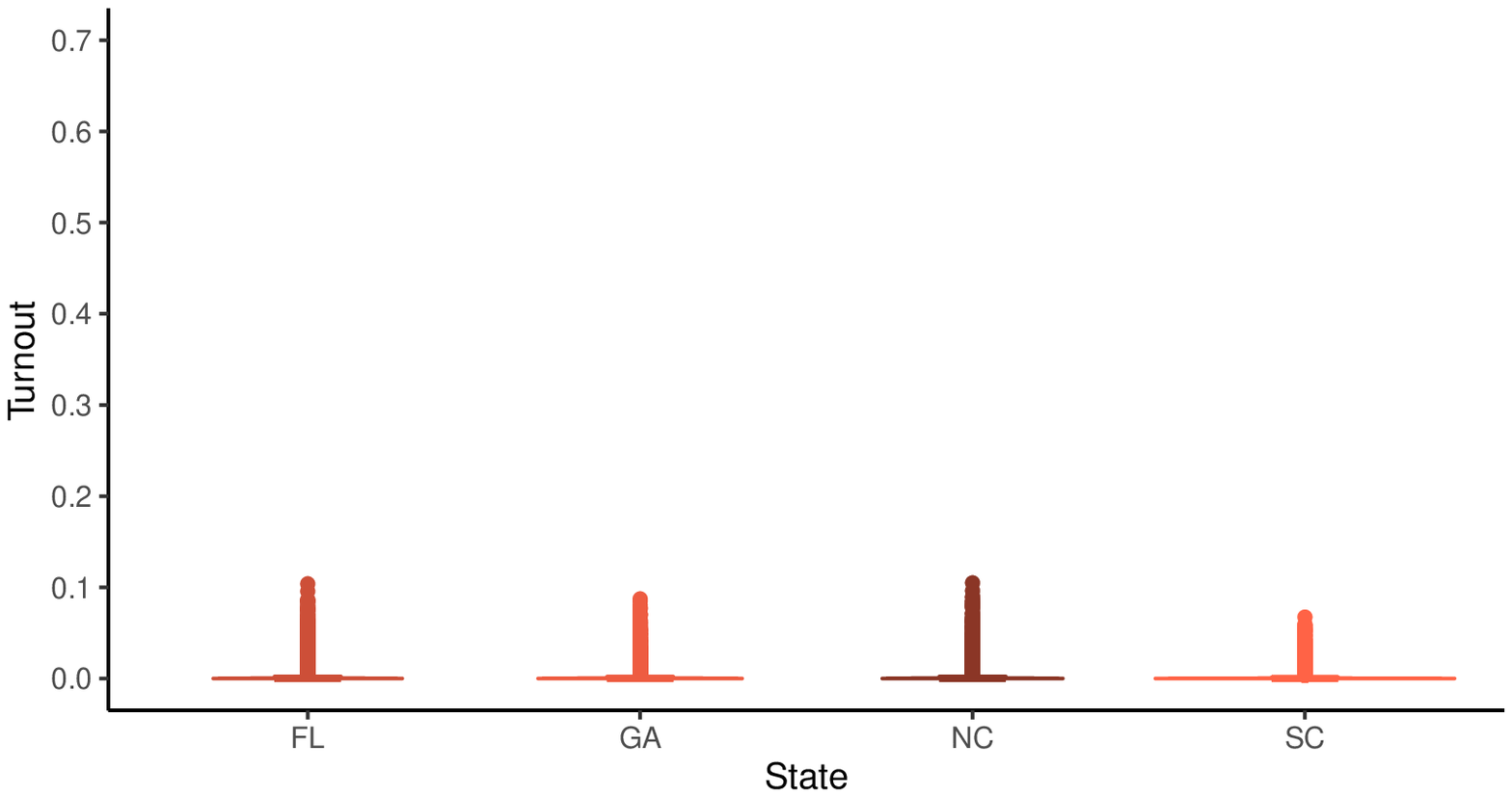}}
} \vspace{10pt}
\subfloat[Predicted turnout among unit nonrespondents for MD-R model.]{
\makebox{\includegraphics[width=0.45\linewidth,height=3.7cm]{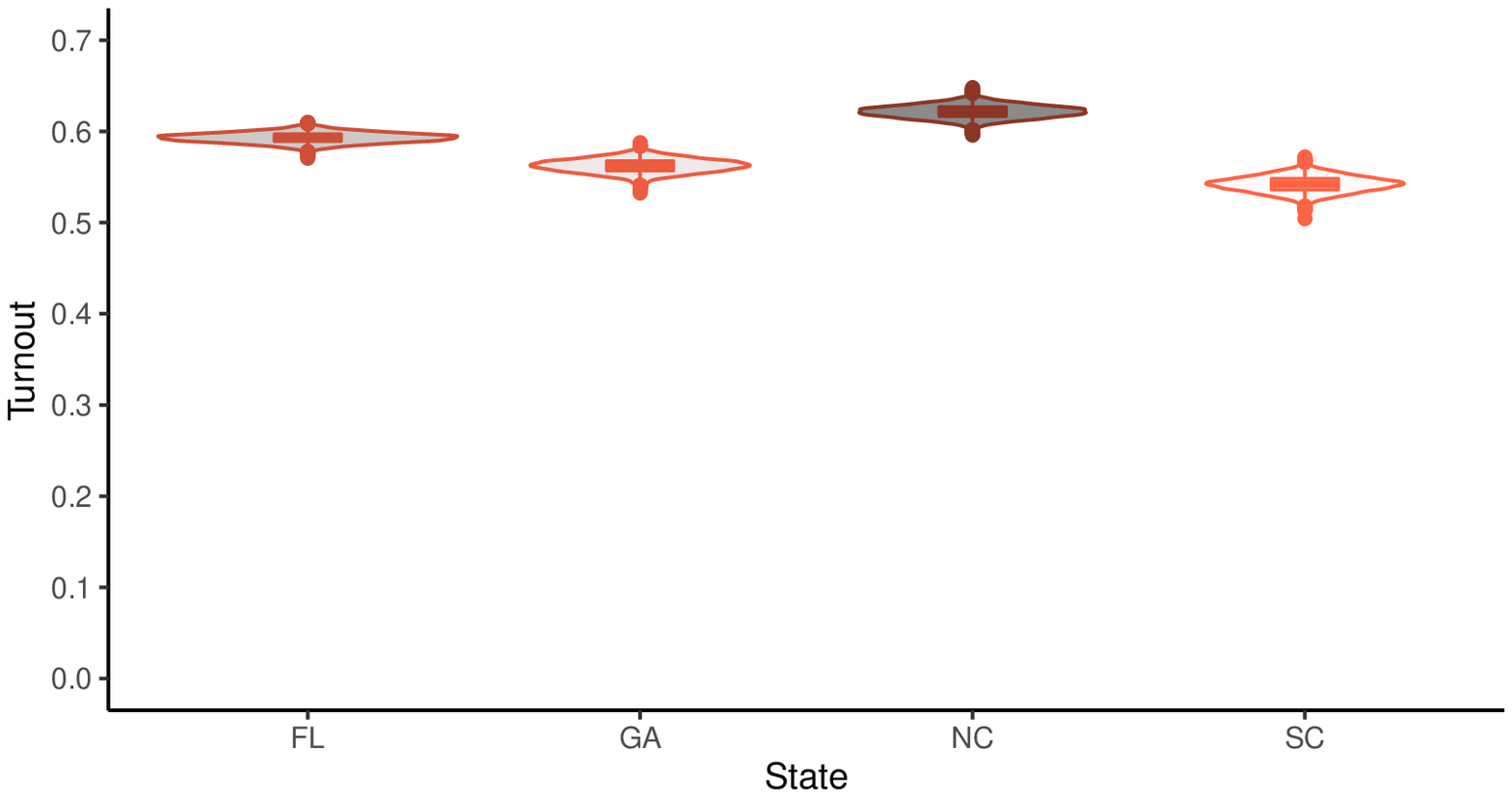}}
}\\
\subfloat[Predicted turnout among item nonrespondents for MD-U model.]{
\makebox{\includegraphics[width=0.45\linewidth,height=3.75cm]{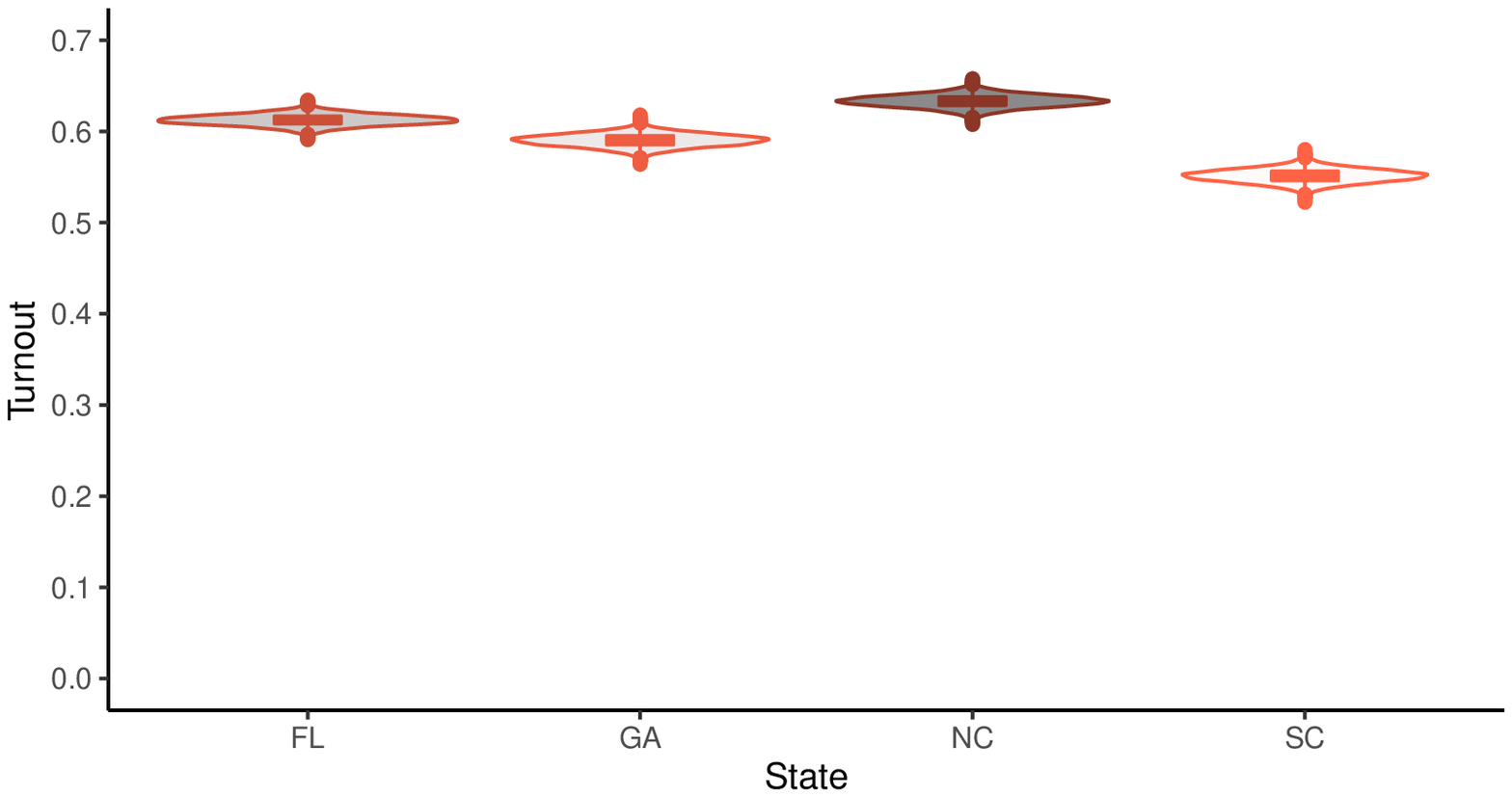}}
} \vspace{10pt}
\subfloat[Predicted turnout among unit nonrespondents for MD-U model.]{
\makebox{\includegraphics[width=0.45\linewidth,height=3.75cm]{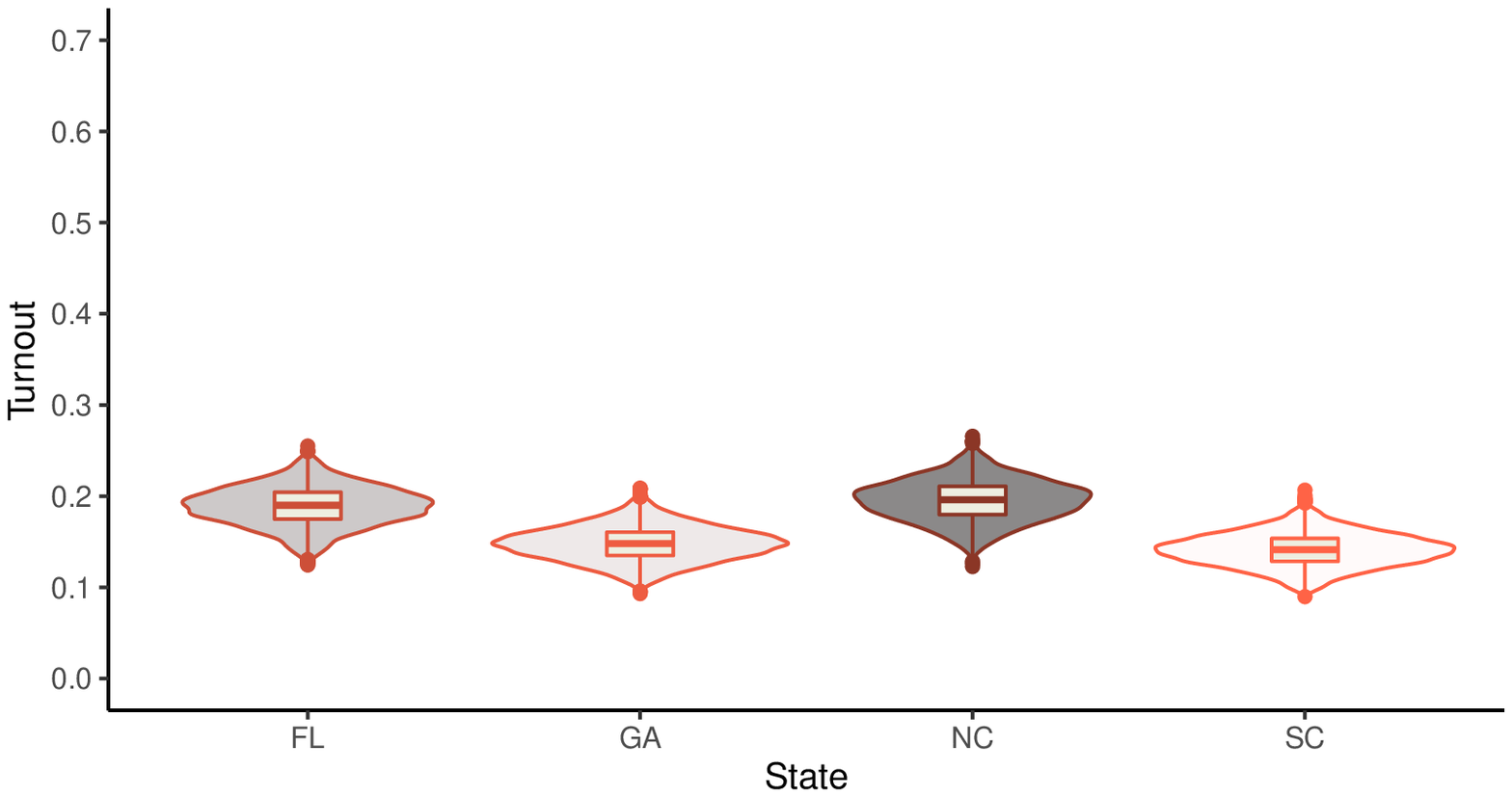}}
}
\caption{\label{results1}Posterior predicted turnout among item and unit nonrespondents.}
\end{figure}

To visualize the implications of the different selection model assumptions, we examine the posterior predictive distributions for turnout for those who did not respond to the survey.  These are displayed in  Figure \ref{results1}.
Across the states, MD-R predicts almost all the  item nonrespondents to be non-voters and more than half of the unit nonrespondents to be voters. MD-U, on the other hand, predicts less than one-quarter of the unit nonrespondents to be non-voters and more than half of the item nonrespondents to be voters.  These differences arise mainly because (i) the gaps between the population turnout margins and complete case point estimates are very large and (ii) item nonresponse rates are low compared to unit nonresponse rates.  Essentially, for any model based on the MD-AM framework, to make up for the large differences, many nonrespondents must be predicted as nonvoters. Using $V$ as a predictor for item nonresponse encourages MD-R to impute as many item nonrespondents as possible to be nonvoters, which must include nearly all of them to get close to the VEP margin. On the other hand, using $V$ as a predictor for unit nonresponse encourages MD-U to impute more unit nonrespondents as nonvoters than before.
Since the unit nonresponse rates across states are generally larger than the nonresponse rates for vote, MD-U is able to pull the estimated turnout rates further from the complete case estimates, because there are more nonrespondents that can be predicted as nonvoters. This is especially true for subgroups where we would expect both low turnout and survey non-response to be common, e.g., for young voters \citep{brick2009nonresponse}.

\begin{figure}
\centering
\subfloat[Using $V$ in the model for $R^x_V$.\label{postpred:model1}]{
\makebox{\includegraphics[width=0.48\linewidth,height=8cm]{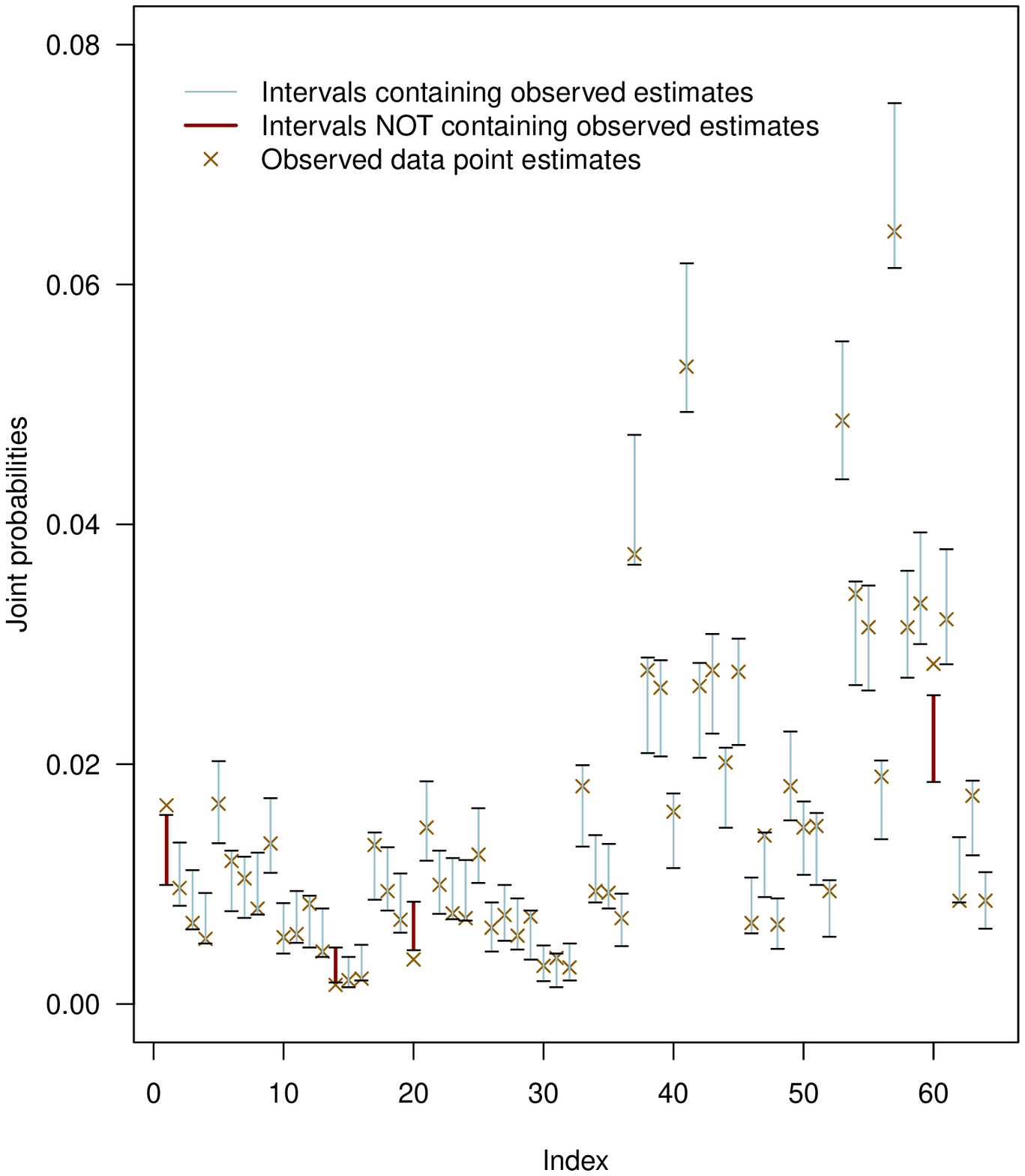}}
}
\subfloat[Using $V$ in the model for $U$.\label{postpred:model2}]{
\makebox{\includegraphics[width=0.48\linewidth,height=8cm]{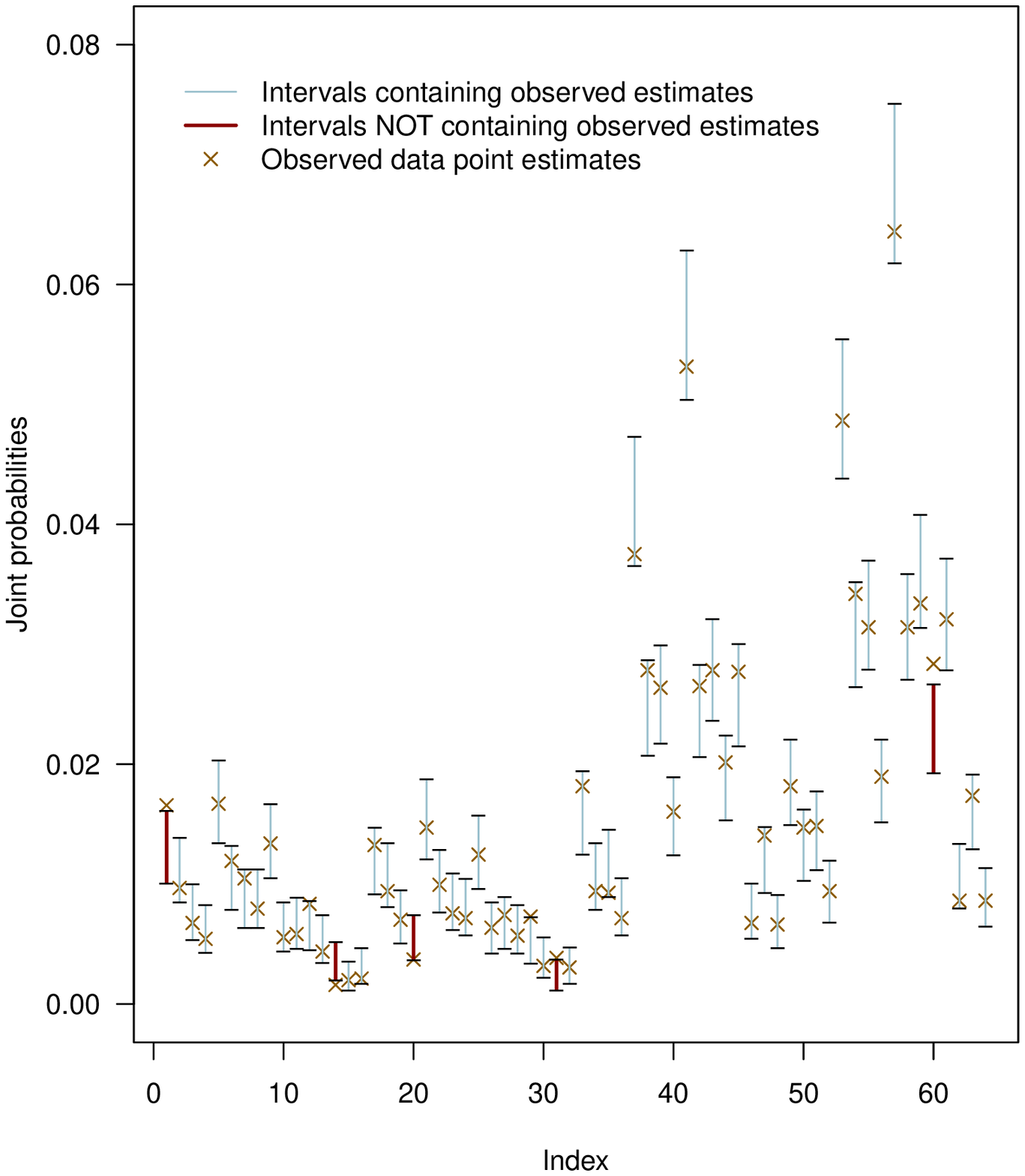}}
}
\caption{\label{postpred}Posterior predictive intervals for all observable joint probabilities from the contingency table for $\mathcal{D}$.}
\end{figure}

An additional way to assess and compare the MD-AM models is to examine how well they fit the observed data.  To do so, we construct 95\% posterior predictive intervals for all 64 four-way observable joint probabilities from the contingency table for $\mathcal{D}$, that is the contingency table for $\{(S_i, G_i, A_i, V_i): U_i=0, R^{y}_{iG} = 0, R^{x}_{iA}=0, R^{x}_{iV}=0\}$, based on all 5,000 posterior samples. As seen in Figure \ref{postpred}, both MD-AM models fit the observed data reasonably well. For MD-R,  approximately 94\% of the predictive intervals contained their corresponding observed data point estimates. The results are nearly identical for MD-U.  Approximately 92\% of the  predictive intervals contained the estimates of the four-way joint probabilities from the observed data. Evidently, both MD-AM models do a good job of capturing the joint relationships in the observed data.

\begin{figure}
\centering
\subfloat[Results for MD-R.\label{results3:model1}]{
\makebox{\includegraphics[width=0.48\linewidth,height=4.5cm]{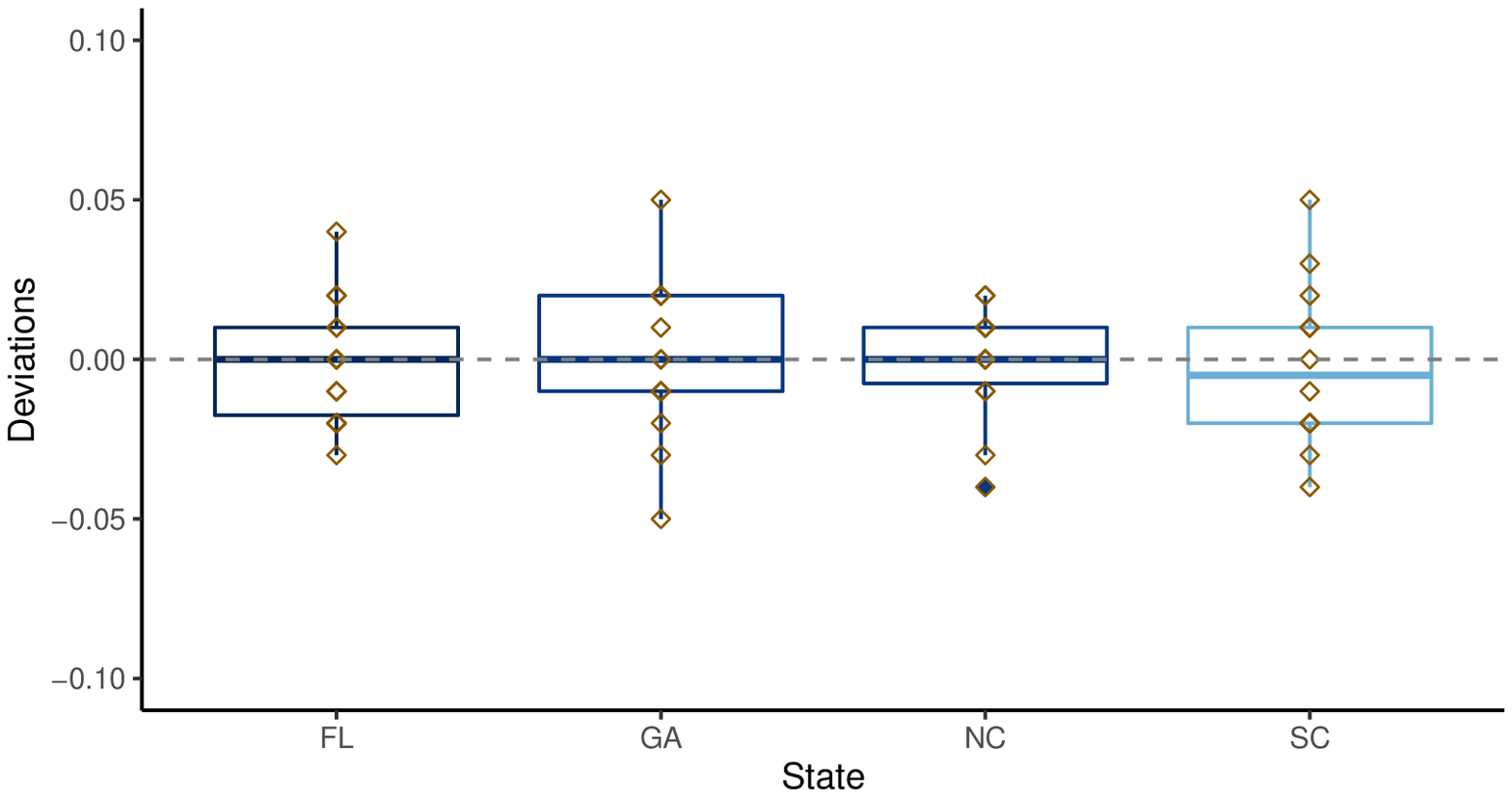}}
}
\subfloat[Results for MD-U.\label{results3:model2}]{
\makebox{\includegraphics[width=0.48\linewidth,height=4.5cm]{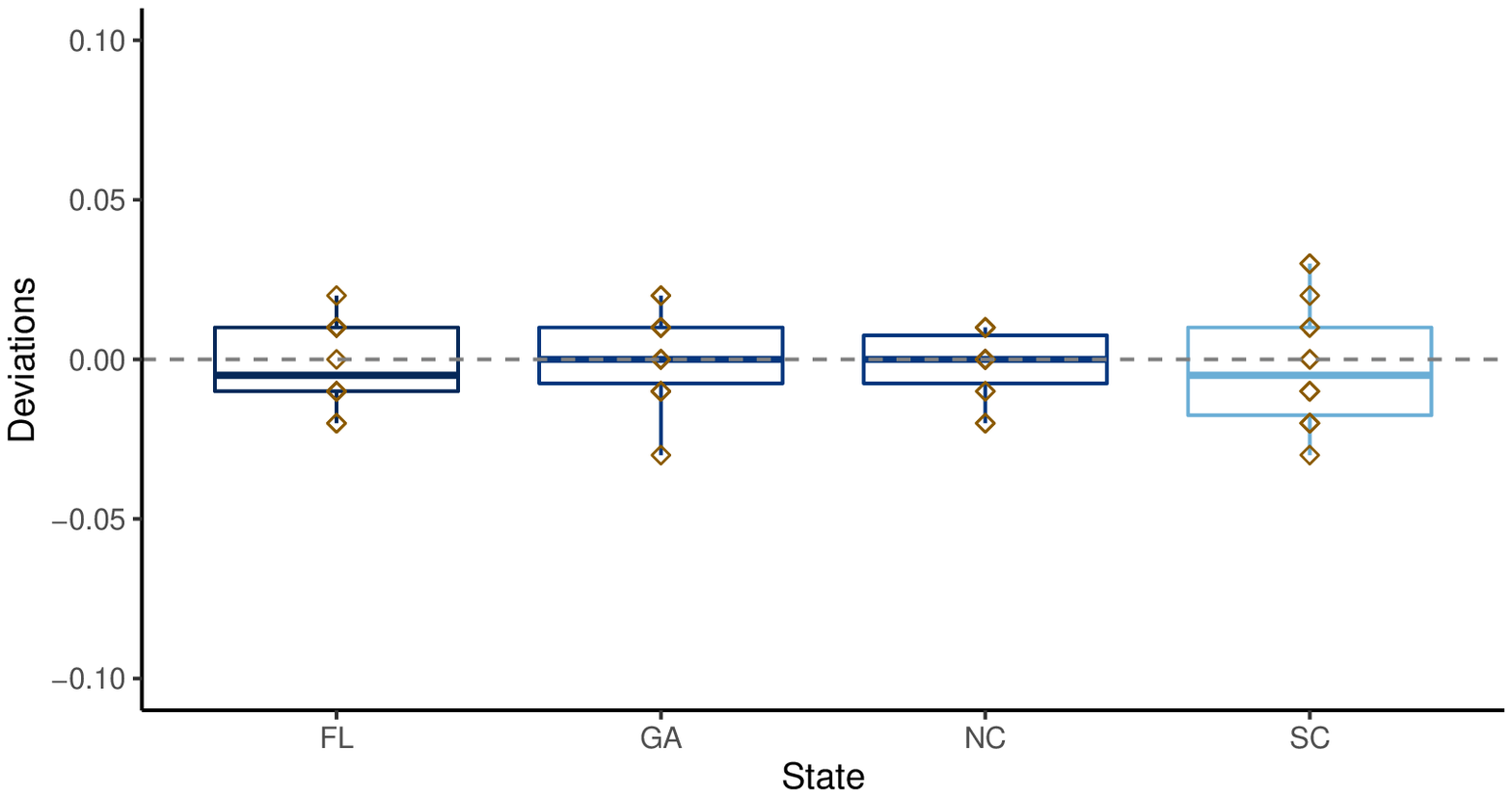}}
}
\caption{\label{results3}Distributions of deviations in MD-AM estimates for voter composition from Catalist estimates.}
\end{figure}

Finally, we validate the MD-AM turnout estimates using voter files in the states as an exogenous benchmark, comparing the posterior predictive demographic composition of \textit{voters} from the models to voter file results.  We rely on voter files compiled by the data firm, Catalist. The Catalist data offer fully observed estimates (with negligible standard error) of the joint probabilities of turnout rates for each subgroup variable we included in our analysis. However, reliable auxiliary marginal information in the Catalist data is available only for those registered to vote, and not for the entire voting eligible population.  Figure \ref{results3} displays the comparisons between the composition of voters in the Catalist data and the MD-AM estimates in each of the four states. Both MD-AM models produce a composition of voters that is very similar to those produced using Catalist's verified voters in these states.

\section{Discussion} \label{discussion}
The MD-AM framework provides a flexible, model-based approach to handle unit and item nonresponse when population-level margins for some survey variables are available from external data sources. Specifically, the framework allows analysts to leverage information on marginal distributions to identify extra parameters in nonresponse models. Analysts can dedicate these extra parameters to the models where flexibility is most beneficial, thereby allowing for nonignorable missing data models and enrichening nonresponse modeling more broadly.

We presented MD-AM models specified as a sequence of parametric regressions.  This can be challenging in practice with large numbers of variables, as the number of possible model specifications can be very large, especially when one considers interaction terms. Model selection is further complicated by identification constraints. Thus, an important future research topic is to incorporate more flexible modeling techniques, such as Bayesian nonparametric methods \citep{si:reiter:hillygus15, si:reiter:hillygus16}, and regression trees \citep{Chipman2010}, in the MD-AM framework.

With respect to the CPS application, we expect that the large discrepancies between the complete case results and the state-wide turnout rates mostly reflect the effects of nonignorable nonresponse. However, they  also could reflect an upward bias from using self-reported turnout data \citep{debell2018turnout}. Some respondents are inclined to say they voted even though they did not, because it is socially desirable to vote. This upward bias has been found to be around 6\% in other election surveys \citep{EnamoradoImai2019,jackman2014does}. One area for future improvement is to deal more carefully with the reporting error in voter turnout. The MD-AM framework could be extended to handle reporting error through a hierarchical specification, as we can add a reporting model to explain how reported values are generated from the true unobserved values.

The Census Bureau recognizes potential over-reporting in how it handles missing turnout values.  For responding households in the CPS, citizens of voting age are counted as nonvoters if they have a response of ``Don't Know,'' ``Refused,'' or ``No Response.''  As explained in the Census Bureau documentation: ``Nonrespondents and people who reported that they did not know if they voted were included in the `did not vote' class because of the general overreporting by respondents in the sample'' \citep{Census2010}.  The Census Bureau uses weighting adjustments for unit nonresponse, calibrating to variables other than the VEP vote totals. Effectively, this ends up making assumptions in the same spirit as those in MD-R; indeed, MD-R ends up imputing most missing turnout items as not voted.  As a result, the CPS and MD-R estimates are closer to one another than the complete-case and MD-R (or MD-U) results; see the online supplement for the CPS results.  However, there are differences.  For example, the state-wide estimates of turnout in NC and SC using the CPS imputations are 69\% and 65\%, respectively, which are quite a bit larger than the corresponding VEP margins of 64.8\% and 56.3\%, suggesting they tend to over-estimate turnout.
In fact, the CPS imputation approach generated the opposite problem in the 2008 election, when official CPS estimates indicated a turnout rate that was slightly lower than that in 2004, despite a historic number of ballots cast in the 2008 presidential election between Barack Obama and John McCain \citep{hur2013coding}. 

Nonetheless, many agencies traditionally use survey weight adjustments for unit nonresponse and imputation approaches for item nonresponse.  Although this is not a model-based approach like MD-AM, agencies still can leverage the MD-AM framework in this paradigm. Using the CPS application as an illustrative example, the agency could choose to impute respondents' missing items using ICIN or MAR models, and adjust respondents' survey weights for unit nonresponse using the margins for state, age, and vote. Alternatively, the agency instead could impute respondents' missing turnout using additive nonignorable modeling, and adjust respondents' survey weights for unit nonresponse using only the known margins for state and age. When agencies do not provide information on numbers of unit nonrespondents, this may be the most sensible choice for secondary data analysts seeking to use auxiliary margins in item nonresponse modeling. If desired, the agency could evaluate predictive distributions and model fit, as we do in Section \ref{illustration}, using model-based MD-AM framework to inform decisions about which uses of the margins seem to offer the most plausible results. Even better,  agencies could make multiple data files available to enable sensitivity analyses, which arguably would enhance current practice.

The framework presented here does not incorporate survey weights in the imputations.  Partly, this is because we do not have the design weights in the CPS for participants---the Census Bureau adjusts their weights for nonresponse and post-stratification---or for unit nonrespondents. Indeed, the CPS documentation does not even provide the number of unit nonrespondents to the voter supplement---as an aside, our work indicates that it would be useful for agencies to release such details. A first step at a method for incorporating survey weights in the MD-AM framework was put forward by \cite{Akande2019} and \citet{akande:reiter:fienbergbook}, who outline a general approach for doing so in stratified sampling. While fully describing this first step is beyond the scope of this article, the basic idea is to add a condition to the MD-AM model that $\hat{t}_{X_k} \sim N(t_{X_k}, V_k)$, where $\hat{t}_{X_k}$ is the Horvitz-Thompson estimator of $t_{X_k}$, the total of $X_k$, and $V_k$ is an agency-specified variance.  This has the effect of ensuring survey-weighted estimates of  totals using the completed data are plausible relative to their known population totals, while allowing for the flexibility of nonresponse modeling.  Extending this or developing other approaches for general complex designs is another important topic for future research.

\section*{Acknowledgement}
This research was supported by grants from the National Science Foundation (SES-1733835 and SES-1131897).

\bibliography{nonresponse.bib}

\end{document}